\documentclass[fleqn]{2017SCGE}
\usepackage{xcolor}
\usepackage{ulem}
\begin{document}

\ensubject{subject}

\ArticleType{Article}
\SpecialTopic{SPECIAL TOPIC: }
\Year{2017}
\Month{}
\Vol{}
\No{}
\DOI{}
\ArtNo{}
\ReceiveDate{}
\AcceptDate{}

\title{Moments of inertia of neutron stars in relativistic mean field theory: the role of the isovector scalar channel}{Moments of inertia of neutron stars in relativistic mean field theory: the role of the isovector scalar channel}

\author[1]{Zhuang Qian}{}
\author[1]{Ruo Yu Xing}{}
\author[1]{Bao Yuan Sun}{sunby@lzu.edu.cn}

\AuthorMark{Bao Yuan Sun}

\AuthorCitation{Qian Z, Xing R Y and Sun B Y}

\address[1]{School of Nuclear Science and Technology, Lanzhou University, 730000 Lanzhou, China}


\abstract{With the inclusion of the isovector scalar channel in the meson-nucleon couplings, taking DD-ME$\delta$ as an effective interaction,
the moments of inertia of neutron stars possessing various stellar masses are studied within the density dependent relativistic mean field (RMF) theory.
The isovector scalar channel contributes to the softening of the neutron-star matter equation of state (EOS) and therefore the reduction of the maximum mass and radius of neutron stars.
Smaller values of the total moment of inertia $I$ and the crustal moment of inertia $\Delta{I}$ are then obtained in DD-ME$\delta$ via numerical procedure in comparison with those in other selected RMF functionals.
In addition, the involvement of the isovector scalar channel lowers the thickness of the neutron star crust and its mass fraction as well. The sensitivity to both the crustal mass and stellar radius causes the crustal moment of inertia to be more obviously reduced than the total one, eventually leading to a suppression on the fraction of crustal moment of inertia $\Delta{I}/I$ in DD-ME$\delta$.
The results indicate the crustal moment of inertia as a more sensitive probe of the neutron-star matter EOS than the total one, and demonstrate that the isovector scalar meson-nucleon couplings in the RMF theory could exert influence over the physics of pulsar glitches.}

\keywords{neutron star, moment of inertia, relativistic mean field theory, isovector scalar channel}

\PACS{
21.60.Jz,~
26.60.-c,~
26.60.Kp,~
26.60.Gj,~
97.60.Gb,~
97.10.Kc
}

\maketitle


\begin{multicols}{2}

\section{Introduction}
As the natural laboratories in the universe for nuclear and particle physics, neutron stars \cite{Lattimerscience536} have generated much effort concentrated on exploring the equation of state of baryonic matter at low temperature and high density \cite{WEBER200794}. Specifically, the observed maximum mass of neutron stars produces a strong constraint on the behavior of EOS at supranuclear density. Besides, as a rotating object its moment of inertia which is one of the crucial bulk properties of neutron stars can be measured in a few years from spin-orbit coupling in double pulsar systems \cite{Lyne2004SCI303.1153, Kramer2009CQG26.073001}. Such a measurement of neutron star moment of inertia would delimit EOS significantly \cite{Lattimer2005ApJ629979} and be used to distinguish neutron stars from quark stars \cite{Yagi2013SCI341.365}.
\Authorfootnote
Thus, precise and massive modern astronomical observations of the moments of inertia, reflecting mass distribution in neutron stars interior, would provide a powerful probe of their internal structure and of the EOS.

Furthermore, the moment of inertia that resides in the crust of neutron stars, the region between the surface and the phase transition boundary separating uniform matter and matter containing nuclei \cite{LATTIMER2007109}, plays an important role in understanding the mechanism of pulsar glitches. The abrupt increase of neutron star rotational frequency, namely the glitches, are well believed to be the result of sudden transfers of angular momentum between the neutron superfluid permeating the inner crust and the rest of the star \cite{LATTIMER2007109, Anderson1975Nature256.25, Link1999PRL83.3362, Andersson2012PRL109.241103, ChamelPRL2013}. Based on the astronomical observation and analysis \cite{Wang2010SCPMA, Espinoza2011MNRAS414.1679}, several theoretical studies contributed to understand its mechanism \cite{Anderson1975Nature256.25, Link1999PRL83.3362,  Andersson2012PRL109.241103, ChamelPRL2013} and these studies were compared to the glitch activities of Vela pulsar \cite{Baym1969NAT224.872}. Recently, the effects of entrainment of superfluid neutrons in the crust were investigated and an serious uncertainty was found in constraining the fraction of crustal moment of inertia \cite{Andersson2012PRL109.241103, ChamelPRL2013, LattimerApJ2013, LiAAstrophysicalJournal2016, ChamelPRD2016, Watanabe2017PRL119.062701}. Assuming the neutron star is rotating slowly, the fractional moment of inertia could be related to the star radius and the pressure $P_t$ at core-crust transition density $\rho_t$ \cite{LATTIMER2007109}. At the same time a correlation between $\rho_t$ and the density slope of nuclear symmetry energy $L$ was found in several studies \cite{Xu2009AJ697.1549, Fattoyev2010PRC82.025810, Ducoin2011PRC83.045810, Moustakidis2012PRC.86.015801, Providencia2014EuroPhyJourA50.44, ZWL}.

During the past decades, the covariant density functional (CDF) theory has proved to be very successful in describing the nuclear structure, nuclear excitation and decay modes \cite{Bouyssy1987PRC36.380, Reinhard1989RPP52.439, Nikolaus1992PhysRevC.46.1757, RING1996193, Zhou2003PRL91.262501, Vretenar2005PR409.101, Meng2006PPNP57.470, Meng2006PRC73.037303, Long2006PLB639.242, Liang2008PRL101.122502, SunRMF1, SunRMF2, Zhao2010PhysRevC.82.054319, Meng2011SCPMA, Hua2012SCPMA, Qu2013SCPMA, Meng2015JPG42.093101, Liang2015PR570.1, SunRHF1, SunRHF2, SunRHF3, SunRHF4, SunRHF5, Guo2016SCPMA, Yao2017SCPMA}. As for nuclear matter, the static as well as the rotating neutron star properties have been studied based on the CDF theory \cite{GlendenningBOOK, boguta1977, boguta1981, SUMIYOSHI1995NPA725, Hofmann2001PhysRevC.64.025804, Ban.S.Fneutronstars, LiuPhysRevC.65.045201, Liu2005, LiuPhysRevC.75.048801, B.Y.SNeutronstar, LongPhysRevC.85.025806, S.W, NBZhangIJMPE2013, QiBinCPL2015, BinQiRAA2016, Dong2016SCPMA, NBZhangCPC2017}. At present, the relativistic mean field (RMF) approach, as a most popular version of CDF theory, has been extended to include the isovector scalar channel in the meson-nucleon couplings, i.e., $\delta$ meson \cite{KUBIS1997191, LiuPhysRevC.65.045201, MenezesPhysRevC.70.058801, DDME-delta, SinghPhysRevC.89.044001}.

Indeed, $\delta$ meson exchange is an essential ingredient of all nucleon-nucleon realistic potentials in the QHD scheme \cite{KUBIS1997191} and the inclusion of $\delta$ meson was also suggested in a relativistic Brueckner theory \cite{SugaharaPRC.55.1211, LenskePRC.57.3099, HofmannPRC.64.034314}. When the $\delta$ meson is introduced, the proton Dirac mass becomes larger than the neutron one in neutron-rich matter with increasing density \cite{KUBIS1997191, Liu2005, S.W}. Such an effective-mass splitting is then treated as an important factor in the stability of drip-line nuclei \cite{HofmannPRC.64.034314} and would affect the nucleon transport properties in heavy ion collisions \cite{LIBA2004563, GAITANOS200424NPA}. Besides, the contribution from the isovector scalar $\delta$ meson within the RMF model become substantial for the nuclear matter with large isospin asymmetry, strongly alternating the density dependence of symmetry energy \cite{LiuPhysRevC.65.045201, Liu2005, LiuPhysRevC.75.048801, S.W}, which then may exert a dramatic impact on the description of neutron star properties. In this work, we will focus on the influence of the isovector scalar coupling channel in the RMF theory on the moments of inertia of neutron stars, including both bulk and crustal ones as well as their ratio, in connection with the physics of pulsar glitches.

\section{Theoretical framework}
The theoretical framework of the RMF approach with the isovector scalar meson-nulceon coupling has been well discussed in the literature. The interested reader is referred to Refs. \cite{LiuPhysRevC.65.045201, LiuPhysRevC.75.048801, DDME-delta, S.W} for more details. Starting from the effective Lagrangian density with the inclusion of isovector scalar $\delta$ mesons, the energy density and the pressure for nuclear matter is obtained from the energy-momentum tensor. Then the EOS used in neutron star core region are obtained under $\beta$-equilibrium condition for homogeneous neutron star matter (neutrons, protons, electrons and muons). While the neutron star crust primarily consists of inhomogeneous nucleonic matter, BPS \cite{BPS1971ApJ170299Baym} and BBP \cite{BBP1971225BAYM} models are adopted for proper description of the EOS at the crust region instead of RMF models. For simplicity, the possible degrees of freedom beyond nucleons, such as hyperons and quarks, are not considered inside the neutron stars. Under the Hartree approximation, the isovector scalar coupling channel itself leads to a positive contribution to the energy density and a negative contribution to the pressure. Moreover, the occurrence of $\delta$ mesons suppresses the kinetic and the isoscalar scalar $\sigma$ potential energies resulting from the alternated effective mass, leading to the EOS of nuclear matter softened considerably.

Given the EOS, the stellar structure could be determined by the Tolman-Oppenheimer-Volkov (TOV) equations and the moment of inertia of neutron stars based on the slowly rotating assumptions are defined in the framework of general relativity \cite{Hartel1967APJ1501005, LATTIMER2007109}
\begin{equation}\label{totalI}
\frac{dI}{dr}=-\frac{2c^2}{3G}r^3\omega(r)\frac{dj(r)}{dr},
\end{equation}
where $j(r)=e^{-(v(r)+\lambda(r))/2}$, with the metric functions $\nu(r)$ and $\lambda(r)$ satisfying
\begin{align}
  \frac{dv(r)}{dr}&=2G\frac{M(r)+4{\pi}r^3p(r)/c^2}{r(r-2GM(r)c^2)},\\
  e^{-\lambda(r)}&=1-\frac{2GM(r)}{rc^2}.
\end{align}
The rotational drag $\omega(r)$ is solved from the equation
\begin{equation}\label{omega}
  \frac{d}{dr}(r^4j(r)\frac{d\omega(r)}{dr})=-4r^3\omega(r)\frac{dj(r)}{dr},
\end{equation}
with the boundary conditions required by the continuity at the stellar surface
\begin{align}\label{omega2}
  \omega(R)&=1-\frac{2GI}{R^3c^2},& j(R)&=1.
\end{align}
Starting from a constant trial value of $\omega$ and $d\omega/dr=0$ at $r=0$, the stellar profile of moment of inertia can therefore be numerically obtained by solving the equations above iteratively together with the TOV equations.

To further calculate the crustal moment of inertia of neutron stars, the stability between nuclei and uniform matter in neutron stars that defines the core-crust interface should be discussed. In general, dynamical method is used to estimate the instability region of neutron star matter at $\beta$ equilibrium as a realistic treatment \cite{Ducoin2007NPA789.403, Xu2009AJ697.1549, Ducoin2011PRC83.045810, Piekarewicz2014PRC90.015803}. With the inclusion of the density gradient and Coulomb terms, the dynamical method takes into account the finite size effects, which could enhance slightly the stability of the uniform neutron star matter and correspondingly reduce the core-crust transition density $\rho_t$ by about $0.005\thicksim0.015~\rm{fm}^{-3}$ \cite{Ducoin2007NPA789.403, Xu2009AJ697.1549}. As a simplification of the dynamical one by ignoring the finite size effects, the thermodynamical method has been adopted in the discussions with a variety of nuclear effective models and microscopic approaches \cite{KubisPhysRevC.76.025801, Ducoin2011PRC83.045810, Atta2014PRC90.035802}. Since we mainly focus on the effects of isovector scalar channel, the thermodynamical method is adopted here. A more realistic dynamical treatment will just lead to a systematical shift of $\rho_t$ for the selected RMF functionals and does not change the following conclusion about the role of $\delta$ meson. Thus, one can introduce a quantity $V_{ther}$,
\begin{equation}\label{Vthermal}
V_{ther}
=2\rho\frac{\partial\varepsilon_b}{\partial\rho}
+\rho^2\frac{\partial^2\varepsilon_b}{\partial\rho^2}
-\rho^2(\frac{\partial^2\varepsilon_b}{\partial\rho\partial\chi_p})^2/(\frac{\partial^2\varepsilon_b}{\partial\chi_p^2})
\end{equation}
Where $\varepsilon_b$ is the binding energy per nucleon, $\chi_p = \rho_p/\rho$ the proton fraction.
The intrinsic stability condition is violated at the core-crust transition density $\rho_t$ where $V_{ther}(\rho_t)=0$.

\section{Results and discussion}
\begin{figure}[H]
\centering
\includegraphics[width=0.48\textwidth]{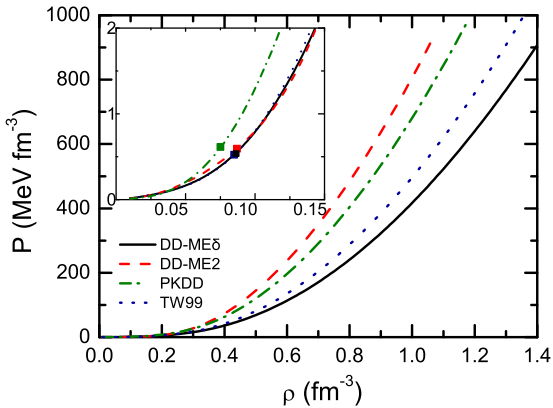}
\caption{The pressure of neutron star matter as a function of the baryonic density. The insert figure displays the pressure at low density where the points on the line indicate the position of the core-crust transition density. The results are calculated by the RMF functionals without the $\delta$ meson coupling, namely TW99, DD-ME2 and PKDD, and the one DD-ME$\delta$ with the $\delta$ coupling.}\label{F:pressure}
\end{figure}

To illustrate how the isovector scalar meson $\delta$ affects the EOS for the nuclear matter, Figure \ref{F:pressure} shows the pressure of neutron star matter calculated by the functional DD-ME$\delta$ as a function of the baryonic density $\rho_b$. The results with the RMF functionals TW99 \cite{TW99}, DD-ME2 \cite{DDME2} and PKDD \cite{PKDD}, in which the $\delta$ meson-nucleon coupling are not included, are included for comparison. It is seen that DD-ME2 provides the stiffest EOS among all functionals, while DD-ME$\delta$ with the isovector scalar channel obtains the softest EOS.

\begin{figure}[H]
\centering
\includegraphics[width=0.48\textwidth]{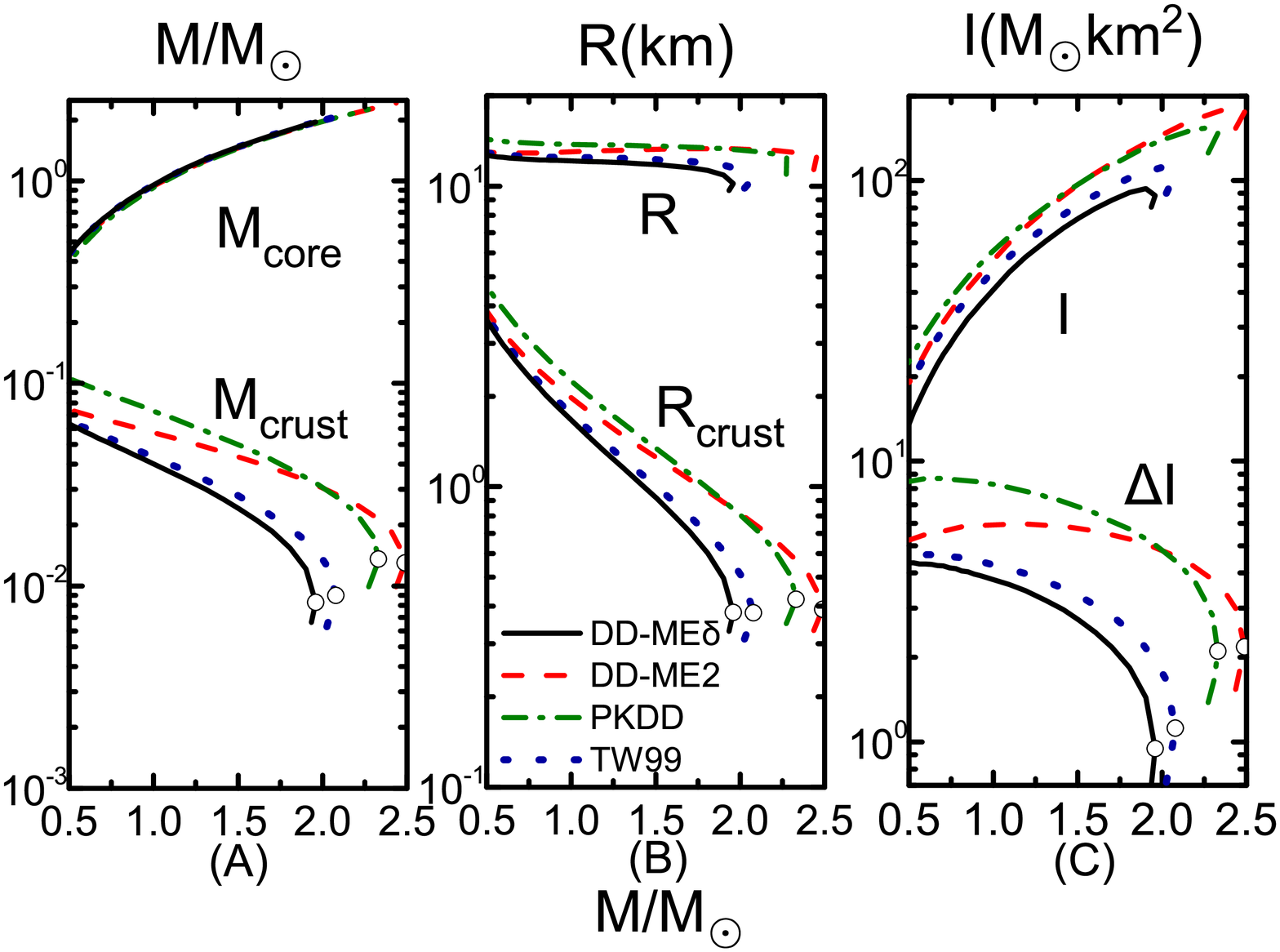}
\caption{Mass, radius and moment of inertia of neutron stars as a function of stellar mass calculated by the selected RMF functionals. Open circles indicate the predicted maximum mass of neutron stars.}
\label{F:M-R-I}
\end{figure}

A stiffer EOS at high densities would provide stronger pressure to sustain the star from collapsing, which in turn leads to a lager maximum mass of neutron stars, as seen in Figure \ref{F:M-R-I}(A) and the detailed mass-radius relation in Ref. \cite{S.W}. Figure \ref{F:M-R-I}(B) displays the neutron star radius as a function of the stellar mass. Because of the effects on the nucleon effective mass and the isoscalar scalar $\sigma$ field, DD-ME$\delta$ provides the softest EOS among the selected functionals, correspondingly the smallest values of the radius prediction.

It is acknowledged that the radius of the neutron star is sensitive to the density dependence of the symmetry energy. In general, a larger value of the slope parameter $L$ implies a stiffer EOS and, hence, a less compact and a more extended neutron star. The correlation between the radius of a 1.4M$_{\odot}$ neutron star $R_{1.4}$ and the density slope $L$ at the saturation density has been illustrated \cite{FattoyevPRC2011}. Here, the values of the slope $L$ and the radius of 1.4M$_{\odot}$ neutron stars with the selected RMF functionals are listed in Table \ref{T:slope}. According to previous studies (e.g. see Ref. \cite{S.W}), PKDD provides the strongest density dependence of symmetry energy around the saturation density and gives, therefore, the largest radius $R_{1.4}$ compared to the other functionals. However, $R_{1.4}$ cannot be uniquely constrained by the symmetry energy but also extremely sensitive to the equation of state at high density \cite{PiekarewiczPRC2001}. From Tab. \ref{T:slope}, DD-ME$\delta$ provides larger value of $L$ than DD-ME2, but shows the smaller value of $R_{1.4}$, demonstrating clearly the crucial role of the stiffness of EOS at high density to the star radius.

\begin{figure}[H]
\centering
\includegraphics[width=0.48\textwidth]{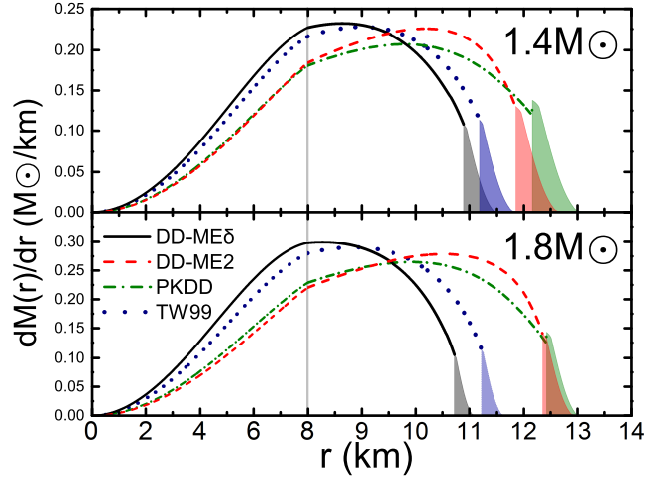}
\caption{The radial dependence of mass distribution $\frac{dM(r)}{dr}=4\pi{r^2}\rho(r)$ in 1.4$\rm{M}_{\odot}$ and 1.8$\rm{M}_{\odot}$ neutron stars. The crust region of neutron star is shown in the shaded area. Note that the scale on the right side of the gray vertical line is 1 km rather than 2 km.}\label{F:add}
\end{figure}

Figure \ref{F:add} displays the radial dependence of mass distribution for 1.4$\rm{M}_{\odot}$ and 1.8$\rm{M}_{\odot}$ neutron stars. Note that due to the low core-crust transition density as shown in Tab. \ref{T:slope}, in the PKDD case, there is an abrupt increase of $dM(r)/dr$ at the core-crust transition boundary in order to satisfy the continuity of pressure in the star. With softened EOS, the density of neutron star matter with DD-ME$\delta$ in the core is enhanced to provide the pressure needed against collapse. Therefore, the neutron star is made to be more compact for a given stellar mass, consequently the crustal thickness narrower as highlighted by the shaded area in Fig. \ref{F:add}. Additionally, such compactness of DD-ME$\delta$ is found to be strengthened as the stellar mass increases.

Aside from the stellar mass, the moment of inertia of neutron stars also depends sensitively on the star radius, approximately proportional to $R^2$. Hence, the properties of neutron star radius and mass reflect the nature of neutron star's moment of inertia to a certain extent. Naturally, as seen in Fig. \ref{F:M-R-I}(C), DD-ME$\delta$ provides the smallest prediction of the moment of inertia, due to the smallest radius displayed in Fig. \ref{F:M-R-I}(B).

\begin{table}[H]
\renewcommand{\arraystretch}{1.5}
\begin{tabular}{ccccc}
\hline
\hline
Interaction         & DD-ME$\delta$ & DD-ME2 & TW99 & PKDD \\
\hline
\hline
$L$ (MeV)&        $52.58$&$51.21 $&$55.31 $&$90.25 $\\
\hline
$R_{1.4}$ (km)&           $12.013$&$13.27$&$12.361$&$13.685$\\
$\rho_{1.4}~(\rm{fm}^{-3})$&        $0.52 $&$0.34 $&$0.47 $&$0.35 $\\
\hline
$\rho_t~(\rm{fm}^{-3})$&        $0.086$&$0.087 $&$0.085 $&$0.075 $\\
$P_t~(\rm{MeV~fm}^{-3})$&        $0.537$&$0.593$&$0.524$&$0.614$\\
\hline
\hline
\end{tabular}
\caption{The density slope parameter of the symmetry energy $L$, the radius $R_{1.4}$ and the central density $\rho_{1.4}$ for 1.4M$_{\odot}$ neutron stars, the density $\rho_t$ and pressure $P_t$ at the core-crust transition boundary calculated by the selected RMF functionals.}
\label{T:slope}
\end{table}

\begin{figure}[H]
\centering
\includegraphics[width=0.48\textwidth]{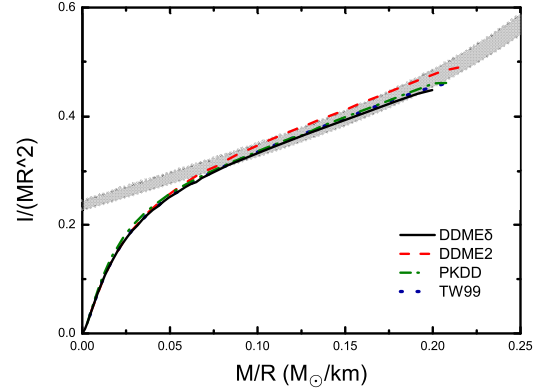}
\caption{The moment of inertia scaled by $MR^2$ as a function of the stellar mass-radius ratio $M/R$ calculated by the selected RMF functionals. The shaded band illustrates the relation of Eq. \eqref{inertia}.}
\label{F:I-mr2}
\end{figure}

Figure \ref{F:I-mr2} displays the moment of inertia scaled by $MR^2$ as a function of the stellar mass-radius ratio $M/R$. It has been suggested that there appears to be a relatively unique relation between $I/MR^2$ and $M/R$ for mass greater than 1.0~M$_{\odot}$  \cite{Lattimer2005ApJ629979}, unless the EOS has an appreciable degree of softening, possibly introduced by hyperons, Bose condensates or self-bound strange quark matter, which is then expressed as
\begin{equation}\label{inertia}
I\simeq(0.237 \pm 0.008)MR^2(1+2.84\beta+18.9\beta^4),
\end{equation}
where $\beta=GM/(Rc^2)$ is the neutron star compactness parameter. All the results of the selected RMF functionals are found to be in good agreement with the approximate relation.

Once the critical density $\rho_t$ of core-crust transition is determined by the thermodynamical condition $V_{ther}(\rho_t)=0$, as seen by the points in the inserted plot of Fig. \ref{F:pressure}, the crustal properties of neutron stars could be calculated numerically. From Tab. \ref{T:slope}, the functional DD-ME$\delta$, due to extra negative contribution of isovector scalar channel to the pressure, provides smaller transition pressure $P_t$ than PKDD and DD-ME2, although still comparable to TW99. The transition pressure $P_t$ prevents the crust from collapsing to the core. Thus, smaller transition pressure would sustain lighter crustal mass for the radius-fixed star. Moreover, smaller stellar radius and crustal thickness would also lead to a reduction in the crustal mass. Figure \ref{F:M-R-I}(B) displays the stellar radius $R$ and the crustal thickness $R_{\rm{crust}}$ where DD-ME$\delta$ provides the smallest value due to the softened EOS at high density. As shown in Fig. \ref{F:add}, the contribution of stellar mass is more obviously dominated by the core, therefore by the stiffness of the equation of state at high densities, especially clear for DD-ME$\delta$. Note that there is a good anti-correlation between the transition density $\rho_t$ and the symmetry energy slope $L$ \cite{Xu2009AJ697.1549, Fattoyev2010PRC82.025810, Ducoin2011PRC83.045810, Moustakidis2012PRC.86.015801, Providencia2014EuroPhyJourA50.44, ZWL}, which has been checked and found to be satisfied well in this work, see Tab. \ref{T:slope} for details.

Hence, it is concluded from the above analysis that both the core-crust transition properties at subsaturation density and the density dependent behavior of EOS at high densities are responsible for determining the crustal mass. The softened EOS at high densities leads to reduced stellar radius and crustal thickness, which, together with lowered core-crust transition pressure, causes the reduction in the crustal mass. Figure \ref{F:M-R-I}(A) and Figure \ref{F:M-R-I}(B) display the crustal mass $M_{\rm{crust}}$ and its thickness $R_{\rm{crust}}$ separately as a function of the stellar mass, in which DD-ME$\delta$ provides the smallest crustal mass and thickness, and correspondingly the moments of inertia for both the total ones $I$ and the crustal ones $\Delta{I}$ are plotted in Fig. \ref{F:M-R-I}(C). Because of the smallest $M_{\rm{crust}}$, $R_{\rm{crust}}$ and $R$ in comparison with the other RMF functionals, DD-ME$\delta$ shows the most suppressed crustal moment of inertia $\Delta{I}$. While the total moment of inertia depends sensitively on the stellar radius for a given stellar mass as illustrated in Eq. \eqref{inertia}, the crustal moments of inertia are not only affected by the stellar radius, but strongly related to the crustal mass. Therefore, the crustal moment of inertia $\Delta{I}$ presents more distinct model dependence than the total one $I$. So it is expected to take $\Delta{I}$ as a more sensitive probe of the neutron-star matter EOS rather than the total one.

\begin{figure}[H]
\centering
\includegraphics[width=0.48\textwidth]{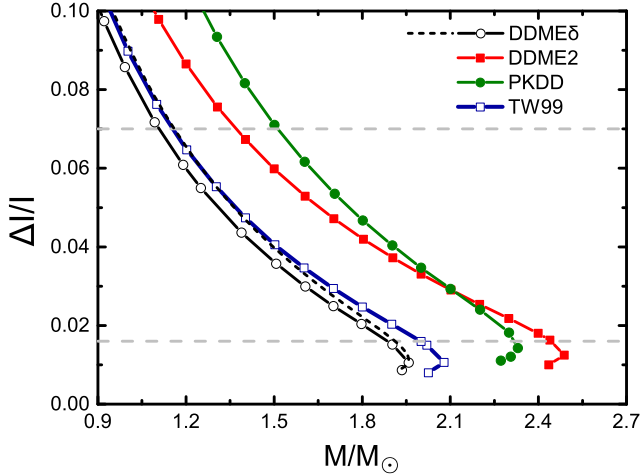}
\caption{The fraction of crustal moment of inertia calculated by the selected RMF functionals. The solid lines are obtained with the slowly rotating assumptions and the dashes line is corresponding results for DD-ME$\delta$ from the expression Eq. \eqref{jiexi}. Two referred constraints of $\Delta{I}/I\geq0.016$ and $\Delta{I}/I\geq0.07$ \cite{Andersson2012PRL109.241103} inferred from the measured glitches in Vela pulsar are presented with the gray horizontal lines respectively.}
\label{F:eta}
\end{figure}

In order to better understand the impact of the $\delta$ meson coupling and the core-crust transition properties on the pulsar glitch phenomenon, it is now appropriated to discuss the fraction of crustal moment of inertia $\Delta{I}/I$. For comparison, an approximation of the ratio can be given by \cite{LATTIMER2007109}
\begin{equation}  \label{jiexi}
\frac{\Delta{I}}{I}\simeq\frac{8{\pi}P_tR^4}{3GM^2}\left[\frac{MR^2}{I}-2\beta\right]e^{-4.8\Delta{R}/R},
\end{equation}
where the moment of inertia $I$ on the rhs. of the expression is just taken simply from Eq. \eqref{inertia}. It is seen that $\Delta{I}/{I}$ is mainly dominated by the transition pressure $P_t$ at the density $\rho_t$ and the radius $R$ of star. Figure \ref{F:eta} shows the calculated fractional moment of inertia $\Delta{I}/I$ of neutron stars based on the core-crust transition pressure taken from Tab. \ref{T:slope} and on the values of $I$ and $\Delta{I}$ resulting from Fig. \ref{F:M-R-I}. For comparison, the approximated calculation adopting Eq. \eqref{jiexi} for $\Delta{I}/I$ is plotted as well for DD-ME$\delta$. For all RMF functionals, $\Delta{I}/I$ decreases monotonously as the stellar mass goes up until to the maximum, and DD-ME$\delta$ provides the smallest value of $\Delta{I}/I$ among the functionals. A little more clear, DD-ME2 and PKDD display larger $\Delta{I}/{I}$ due to stiffer EOS at high densities and higher transition pressure $P_t$ they predict than TW99 and DD-ME$\delta$, while for PKDD the largest slope $L$ performs the extra role as well. For TW99 and DD-ME$\delta$ with similar $P_t$, the difference of $\Delta{I}/{I}$ is mainly caused by their distinct stellar radius and, hence, mainly by the deviation between the EOSs at high densities. Taking into account the effects of the softened EOS becasue of the inclusion of the isovector scalar channel, the crustal properties of neutron stars are more obviously influenced than the neutron star bulk properties.

Then we could discuss about the observational constraints on the fraction of crustal moment of inertia. The observed glitch rates and magnitudes for the Vela pulsar lead to constraint as $\Delta{I}/I\gtrsim0.016$ \cite{Andersson2012PRL109.241103}. Recently, it is argued that due to entrainment of superfluid neutrons in the crust one would have to enlarge the inferred lower limit to $\Delta{I}/I\gtrsim0.07$ and the standard glitch model is then called into questions \cite{Andersson2012PRL109.241103, ChamelPRL2013}. The maximum allowed neutron star mass under both requirements are shown in Figure \ref{F:eta}. It is seen that the functional with $\delta$ meson provides lower maximum allowed mass under both criteria. In spite of the low constraint on maximum mass with entrainment, it is argued that, if take pairing into account explicitly in the calculations of the effects of band structure on the neutron superfluid density in the crust of neutron stars, the standard models of glitches based on neutron superfluidity in the crust can not be ruled out yet \cite{Watanabe2017PRL119.062701}.

\section{Summary}
In summary, the moment of inertia and the fraction of crustal moment of inertia of slowly rotating neutron stars, especially the influence of the isovector scalar $\delta$ meson-nucleon coupling, have been studied within the density dependent RMF theory. It is found that the inclusion of $\delta$ meson channel would soften the EOS and the DD-ME$\delta$ provides the smallest stellar radius and maximum mass among the selected RMF functionals. In spite of the correlation with the density slope of symmetry energy at saturation density, the reduction of stellar radius with DD-ME$\delta$ results dominantly from the softened EOS at high density and a more compact neutron star is therefore obtained. Due to smaller stellar radius, the moment of inertia of neutron star is reduced when $\delta$ meson is included.

Furthermore, the crustal properties of neutron star with the inclusion of $\delta$ meson in the RMF functional are discussed. Note that the thermodynamical method is used as a simplification of dynamical method in determining the core-crust transition properties. The dynamical method which includes density gradient and Coulomb terms should be studied in further discussion. The reduced crustal thickness together with transition pressure $P_t$ in DD-ME$\delta$ require a lighter crustal mass in order to resist the collapse of the crust into the core. Thus, the stellar mass is more obviously dominated by the core for DD-ME$\delta$. Smaller value of crustal mass and thickness were then obtained mainly because of the softened EOS particularly at high densities. It has been shown that the crustal moment of inertia $\Delta{I}$, due to its dependence on both the crustal mass and stellar radius, is more obviously reduced than the total moment of inertia in DD-ME$\delta$. The fraction of crustal moment of inertia $\Delta{I}/I$ is therefore suppressed and DD-ME$\delta$ provides the smallest $\Delta{I}/I$ among the selected RMF functionals.

For Vela pulsar, a maximum stellar mass at around $1.9~\rm{M}_{\odot}$ is then predicted within the DD-ME$\delta$ functional under the constraint of $\Delta{I}/I\gtrsim0.016$. Note that, $\Delta I/I$ cannot be dominated only by the core-crust transition pressure $P_t$ determined at low density of EOS, but also influenced by the radius and crustal mass of star, which could be sensitive to the density slope $L$ of symmetry energy as well as the density dependence of EOS at high density. The crustal moment of inertia $\Delta{I}$ is then expected as a more sensitive probe into the internal structure and the EOS of neutron stars rather than the total one $I$. With more sophisticated theoretical studies on the mechanism of glitches and more plentiful and precise data of pulsars such as  from ``FAST" project \cite{FAST.SC, FAST2011, FAST2016}, the properties of EOS of nuclear matter at various densities and isospin asymmetries could be better understood.
\Acknowledgements{This work is partly supported by the National Natural Science Foundation of China (Grant No. 11375076) and the Fundamental Research Funds for the Central Universities (Grant No. lzujbky-2016-30).}




\end{multicols}
\end{document}